# Development of a new generation of micropattern gaseous detectors for high energy physics, astrophysics and environmental applications

V. Peskov[a1], A. Di Mauro[a], P. Fonte[b], P. Martinengo[a], E. Nappi[c], R. Oliveira[a], P. Pietropaolo[d], P. Picchi[e]

[a]CERN, Geneva, Switzerland
[b]LIP/ISEC Coimbra, Portugal
[c]INFN Bari, Italy
[d]INFN Padova, Italy
[e]INFN Frascati, Italy

**Abstract**

We have developed a cost effective technology for manufacturing various layouts of micropattern gaseous detectors for a wide range of applications. Such devices feature resistive electrodes interfaced to a network of thin readout strips/electrodes. The following three examples of such innovative designs and their applications will be presented: a prototype of a novel double-phase LAr detector with a CsI photocathode immersed inside the LAr, a CsI-RICH detector prototype for ALICE upgrade and GEM-like sensors for environmental safety/security applications.

1. Introduction

In the last two decades, very fast developments have taken place in the field of gaseous detectors for photons and charged particles. Traditional gaseous detectors: wire–type and parallel plate-type (RPCs), which have widely been used in high energy and astrophysics experiments, have serious competitors: Micropattern Gaseous Detectors (MPGDs). However, one problem operating MPGDs is that without special precautions these novel devices can be easily destroyed by sparks, which may occur during the experiments. There are several methods for protecting micropattern detectors and front end electronics from spark-driven damage: segmentation of electrodes, protective diodes, etc. These methods have successfully been implemented in the case of Gas Electron Multipliers (GEMs) and in small-area Micromesh Gaseous Structure (MICROMEGAS).
 An alternative approach, which is becoming more and more popular inside the CERN-RD51 collaboration (see for example [1]), is the use of resistive electrodes. The first micropattern detector with resistive electrodes was a thick GEM [2], and subsequently this approach has also been applied to other detectors: MICROMEGAS [3] and CAT ("Compteur a Trou"[4]. This concept triggered a sequence of similar developments, which are nowadays being followed not only by our team, but by several other groups in the framework of RD51 collaboration (see [1, 5-9] and references therein).

---
[1] Corresponding author: vladimir.peskov@cern.ch

In the last couple of years, our team has pursued a new approach: the use of resistive electrodes segmented into strips with an array of readout strips located under the resistive grid manufactured using a multilayer printed circuit technology. These detectors have several important advantages, for example they are more suitable for large-area detectors and show better features for position measurements. Various designs of such detectors have been developed and successfully tested: resistive microstrip detector [7], resistive microhole-microstrip [7], resistive microgap- microstrip [10] etc.

In this paper we will briefly review the main achievements in this new direction and highlight recent, yet unpublished results/developments.

## 2. Resistive microdot-microhole detector for innovative design of a noble liquid time projection chamber.

The standard dual phase detector is basically a noble liquid Time Projection Chamber (TPC) exploiting a two-phase (liquid-gas) medium in one common cell (see for example [11]). Such TPCs operate as follows: any interaction inside the liquid volume creates an ionization track (containing $n_0$ primary electrons) and a fraction of them $n_0\eta_r$ will recombine with positive ions and produce a burst of primary scintillation light, which is subsequently detected by an array of photomultiplier tubes (PMTs) surrounding this volume. A fraction of electrons, $n_0\eta_d$, will escape recombination and will drift towards the liquid-gas border and will reach it after a time $t_d$. In a strong enough electric field applied across the liquid-gas interface, these electrons can be extracted into the gas phase and then be detected, for example with the help of a scintillation chamber made of two parallel-meshes, where the drifting electrons produce a flash of secondary scintillation light, proportional to $n_0\eta_d$. By measuring the ratio of this scintillation light, one can determine the nature of the interactions and select desirable events. To detect the weak primary scintillation burst, one has to use a considerable number of PMTs operating in coincidence mode which affects the costs for such a device.

The goal of our recent studies was to replace the costly PMTs by a metallic plate coated with a photosensitive CsI layer immersed inside the liquid. Earlier it was discovered that such a plate immersed inside the noble liquids acts as a highly efficient photocathode having high quantum efficiency comparable to that of the PMTs [12, 13]. However, this simple and attractive concept was never implemented in any practical device.

Recently we built and tested the first simplified prototype of such a dual-phase detector exploiting a CsI photocathode inside liquid Ar (LAr) - (see Fig. 1 and ([9] for more details). It was housed in an ultra-high-vacuum cryostat equipped with a continuously working recirculation/purification system, allowing a lifetime of several milliseconds to be achieved for electrons inside LAr. As in the case of any other standard dual-phase detector, two parallel meshes separated by a 1cm gap were placed above the liquid volume acting as a gas scintillation chamber. A PMT ETL9357, was placed a few cm above the scintillation chamber. To convert the UV scintillation radiation (~128 nm) into visible wavelengths the PMT window was coated with a tetrapheny butadiene light shifting layer. In preliminary tests, described in [9], ionization inside the LAr was produced by a point-like alpha source $^{241}$Am deposited on a needle-shaped surface. In the present tests a compact $^{55}$Fe source was used as well. The CsI photocathode immersed in the LAr, was a stainless steel disc of 10 cm diameter coated with a 0.4 μm thick CsI layer. The voltage applied between the meshes was typically 1.5 kV and the typical voltage applied to the CsI photocathode was -3 kV.

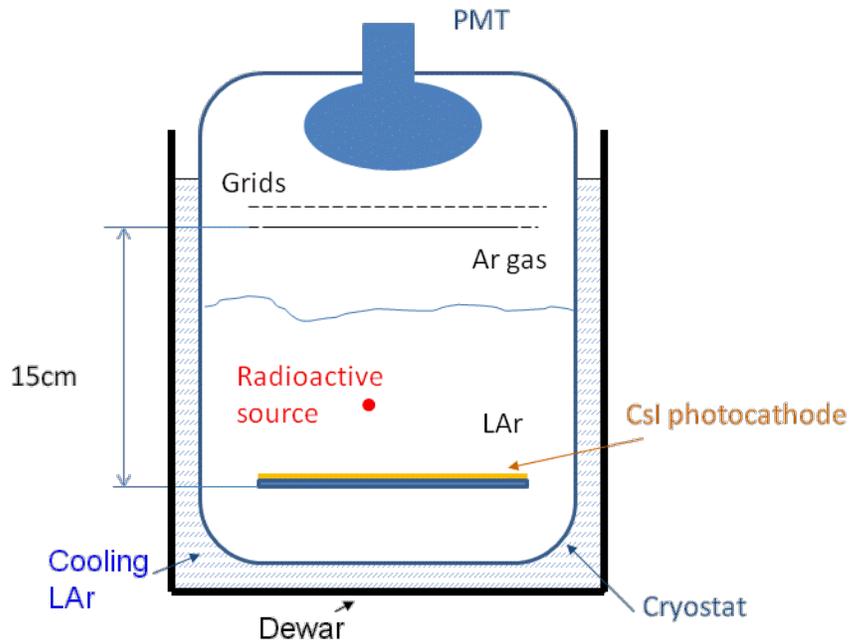

Fig. 1. Schematic drawing of a dual phase detector with a CsI photocathode immersed inside the LAr. As radioactive source either $^{241}$Am or $^{55}$Fe was used.

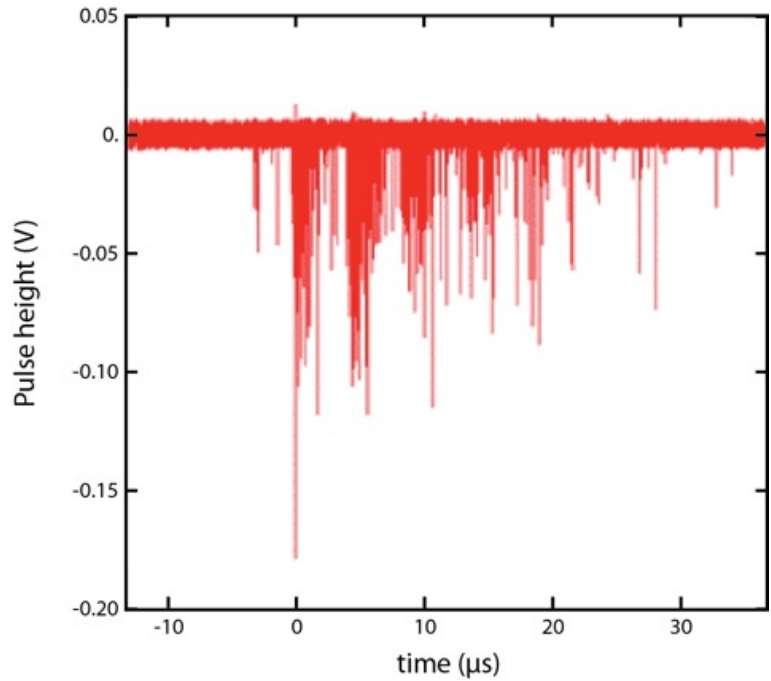

Fig. 2. Typical oscillogram of the PMT signals delivered by the dual phase LAr detector (a single event). The first pulse is the primary scintillation light produced by alpha particles in LAr and the second pulse is the secondary scintillation light produced by primary electrons between the meshes. The other signals are due to the photoelectrons extracted from the CsI photocathode.

As was reported in [9], scintillation light from radioactive sources produces enough primary electrons from the CsI photocathode to be reliably detected. However, the secondary scintillation light, generated between the parallel meshes created undesirable feedback pulses (Fig. 2). These feedback pulses appear because both the primary and the secondary light extract electrons from the CsI photocathode, which then drift to the scintillation chamber and produce other bursts of secondary light. Using a dedicated analysis program we calculated the area under each peak (Fig. 3), in order to obtain a numerical evaluation of the feedback effect. From this data and also taking into account the geometry of the test set-up, we calculated the quantum efficiency of the CsI photocathode to be about 14% for a photon wavelength of 128 nm. Thus we independently confirmed the result obtained in [12-13], that a CsI photocathode immersed inside LAr has quantum efficiency comparable to commercial PMTs. Note that initially the quantum efficiency of the CsI photocathode changes with time till it reaches a saturated value (Fig. 4). This is attributed to the cleaning up process of the CsI photocathode as well as the LAr itself.

In our earlier work [7] to suppress the feedback we suggested using a light /charge multiplication structure that is geometrically shielded from the CsI photocathode. The first successful tests were done with a resistive microhole-microstrip strip plate as a shielding structure [7]. However, considerably higher gains at cryogenic temperatures were later achieved with the microdot-microhole detector schematically shown in Fig. 5. Its principle of operation is similar to the microstrip-microhole detector with metallic electrodes [14]; its manufacturing procedure is described in [9].

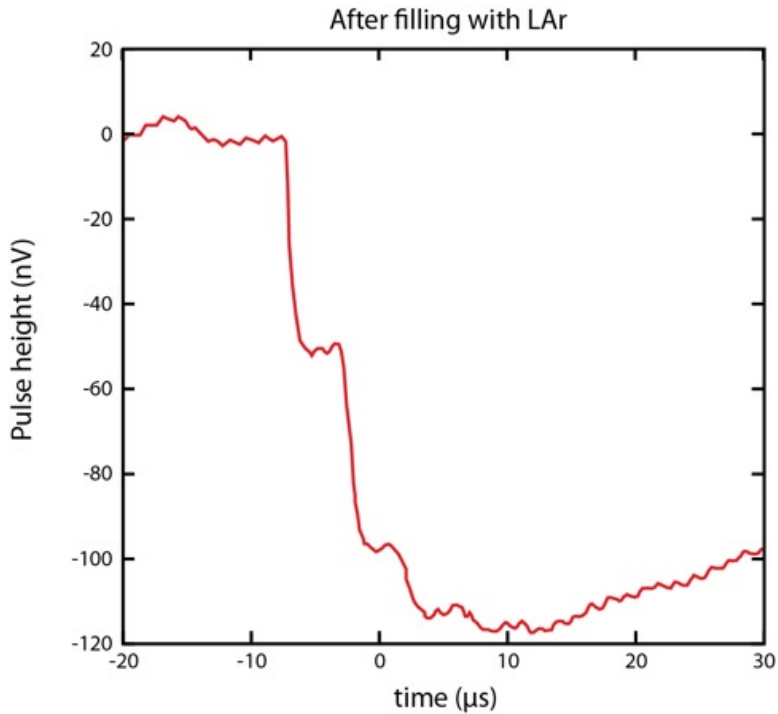

Fig. 3. Integrated signals (200 independent events) for the dual phase LAr detector equipped with the CsI photocathode.

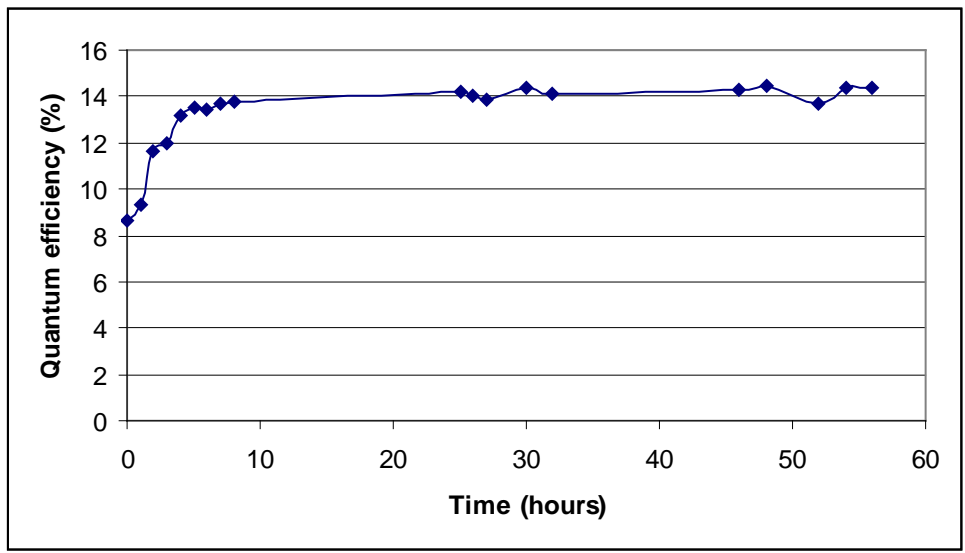

Fig. 4. Time variation of the quantum efficiency of the CsI photocathode immersed inside the LAr.

Fig. 6 shows the gain of this detector as a function of the overall voltage $V_{ov} = V_{bc} + V_{ca}$, (where $V_{bc}$ is the voltage applied between the back-plane and the cathode strips and $V_{ca}$ is the voltage applied between the cathode strips and the anode dots), measured in various gases and at various temperatures using $^{241}$Am and $^{55}$Fe sources. These measurements reveal that with our amplification structure the same gas gain were achieved in pure Ar at room temperature as in the quenched gases clearly indicating that the high gains in pure Ar were possible due to the efficient suppression of feedback from the CsI photocathode (geometric shielding).It can be also concluded that , at room temperature the maximum achievable gain was ~$3x10^4$and this is 3-10 times higher than that achieved with any other avalanche gaseous detectors operating in pure Ar.

It is interesting to note that at critical total charges in the avalanche $Q_{crit}$, which depend on the gas density, self quenched streamers appear (starred data points).

After replacing the parallel-mesh structure by the microdot–microhole detector, the feedback pulses were no longer observed in Ar even at the maximum achievable gains, proving that the light from the avalanche regions (anode dots) is geometrically well shielded with respect to the CsI photocathode.

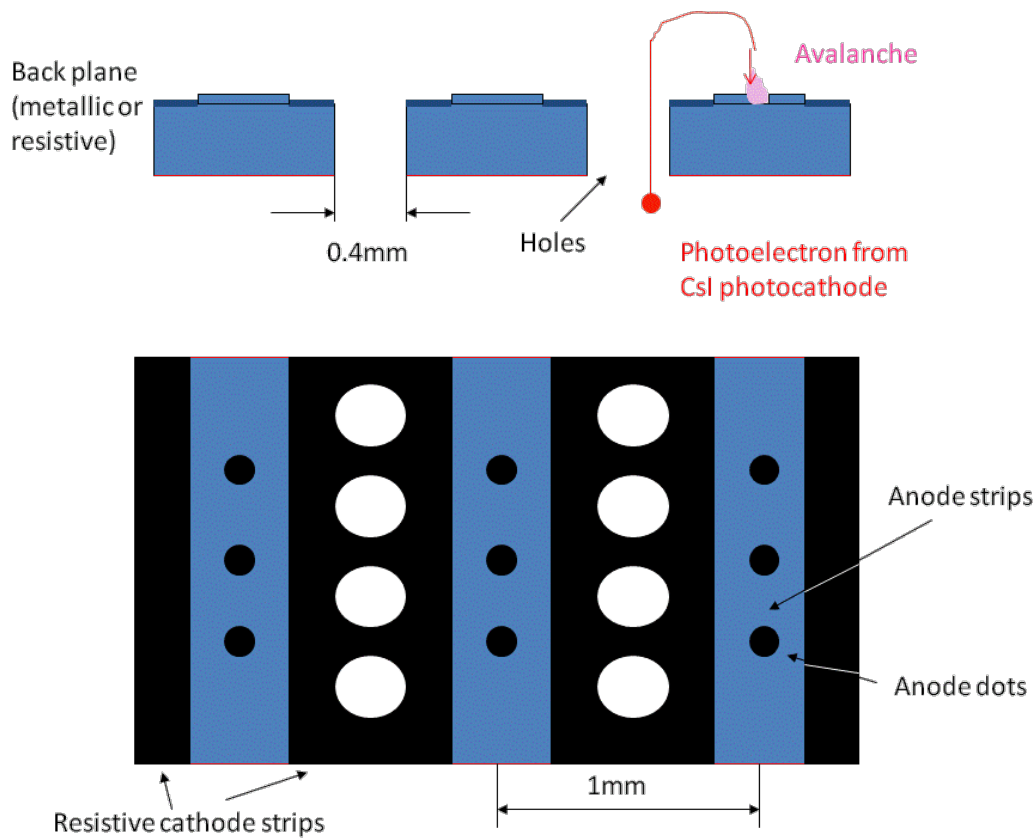

Fig.5. Schematic drawing of a microhole-microdot detector. Photoelectrons extracted from the CsI photocathode drift through the holes towards the anode dots where they experience avalanche multiplication. Because of this geometry, the avalanche light cannot reach the CsI photocathode and create feedback.

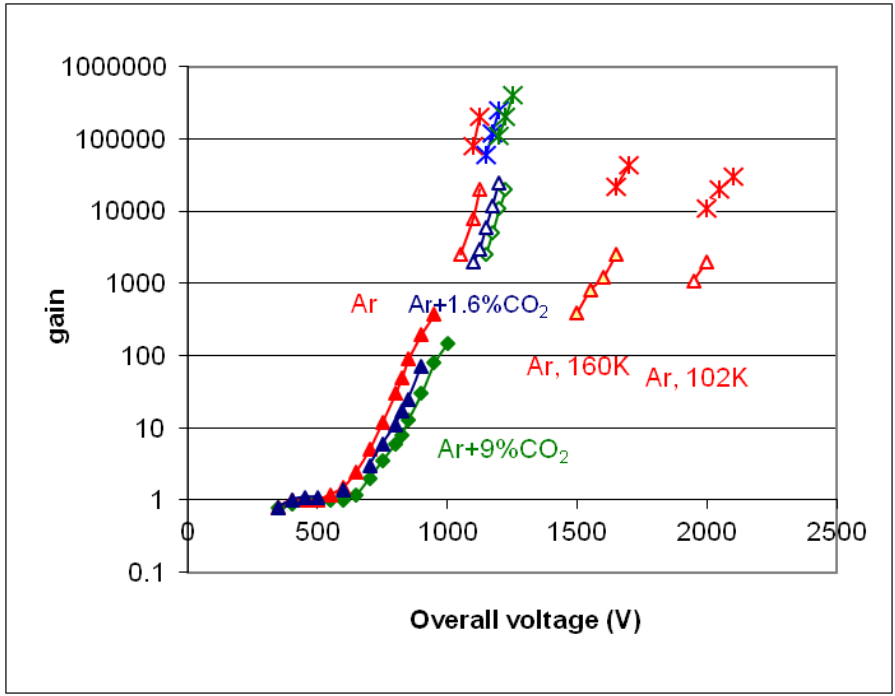

Fig.6. Gas gain curves measured in Ar and Ar+$CO_2$ mixtures at room temperature and in Ar at cryogenic temperatures. Filled symbols refer to measurements with alpha particles

### 3. CsI coated resistive strip GEM combined microstrip detector for RICH applications detector

A couple of years ago, in the framework of the ALICE RICH upgrade [15], our team developed a large-area RICH detector prototype based on CsI coated resistive strip GEMs and succeeded in detecting Cherenkov rings produced by beam particles crossing a liquid or a solid radiator[15-16]. Recently an even simpler RICH prototype based on a resistive microstrip counter (RMSGC) combined with a resistive GEM, the top surface of which was coated with a CsI layer, was developed and tested (see Fig.7 and [7]). This detector could operate at overall gas gains of up to $10^6$ without any sparking or feedback problems. These features should be compared to those of the current ALICE RICH multiwire chamber (MWPC) combined with a CsI photocathode, which can operate without feedback pulses only at gas gains below $10^4$. The reason why GEM-type structures can operate at considerably higher gains is the same as in the case of the cryogenic TPC describe above: the CsI photocathode evaporated on the top of the GEM is geometrically shielded from the avalanche light produced in the holes and/or near microstrips. However, this approach also has a drawback: the number of detected Cherenkov photons is always whereas the open symbols with $^{55}$Fe. Lines with stars show the gas gains in self-quenched streamer mode at room temperature and at 160 and 102 K.

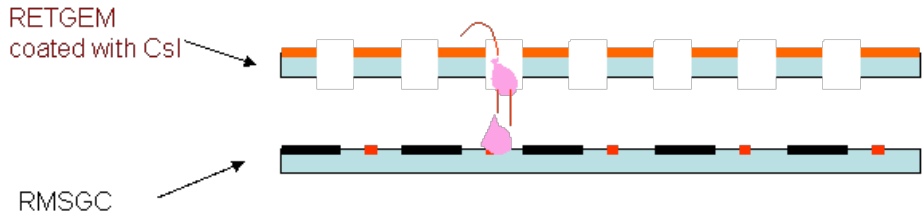

Fig.7. Schematic drawing of the RMSGC combined with CsI-coated RETGEM.

~30% less than in the case of CsI-MWPCs because of the holes in the GEM layers that reduce the effective area of the detector by approximately 25%. To make the resistive microstrip counter combined with the resistive GEM a competitive device with respect CsI-MWPCs, the CsI photocathode was exposed to ethylferrocene (EF) vapors, which form an adsorbed layer on its surface and enhance the quantum efficiency (QE) roughly by 25% (Fig. 8), thus almost compensating for losses due to the holes. The physics behind this enhancement is described in [18, 19].
Note that this method cannot be easily implemented in CsI-MWPCs for the correspondingly increased feedback effect and also because of the undesirable contribution from the EF ionization (our GEMs operate at zero drift field and the contribution from the EF ionization was negligible).

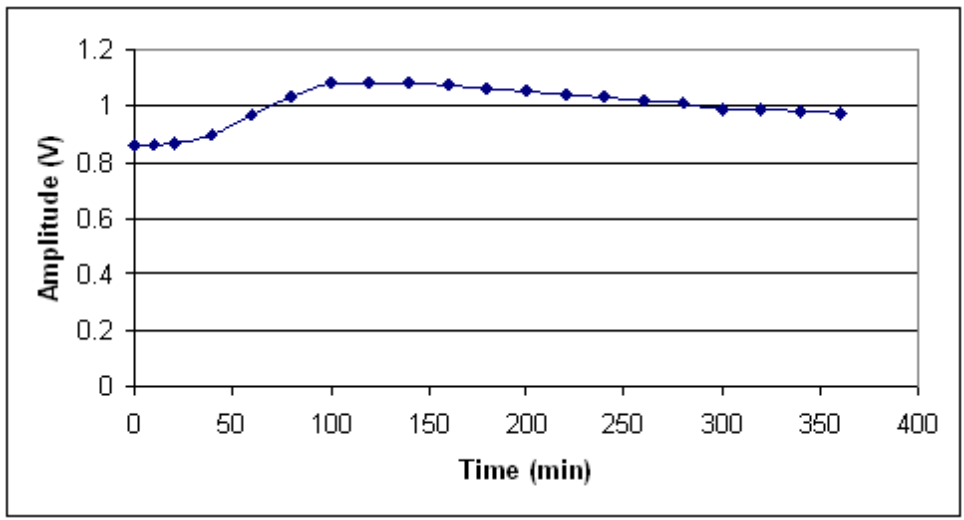

Fig.8. Mean signal amplitude vs. time produced from a resistive microstrip detector combined with a CsI-coated GEM and irradiated by light from a pulsed $H_2$ lamp (160 nm). At t=30 min, EF vapors were introduced into the gas chamber producing the CsI QE enhancement of approximately 25%

## 4. Wall-less resistive GEMs for environmental and safety applications

### 4a. Rn detectors

In previous works [20, 21], we demonstrated that resistive-strip GEMs can operate at high gas gains in ambient air and be used as an efficient detector of Rn. Unfortunately, at air humidity above 30%, a leakage current appears across the inner walls of the GEM holes making the detector noisy and dropping its sensitivity. For this reason, for the detection of Rn single-wire detectors and MWPCs were later developed [22], which had a specially shaped of the dielectric interface between the anode wires and the cathode preventing the leakage current from appearing. Following the same idea, we have developed recently a special GEM–like detector capable of operating in 100% humid air (Fig. 9). This detector consists of two resistive plates with holes supported by a few specially shaped spacers located far away from the holes (we call it a"wall-less GEM").The holes are carefully aligned allowing the formation of an electric field very similar to the one in a standard GEM. Such a structure could operate without spurious pulses at gas gains of $10^3$ in 100% humid air and detect Rn with a sensitivity equal to the best commercial Rn sensors whereas the estimated cost of this detector is at least 10 times smaller. Moreover, the main advantage of this detector is its ability to detect variations in Rn concentration ten times faster than commercial detectors. This was achieved by the fast removal (with the help of a special replaceable drift electrode) of Rn progeny from the detector fiducial volume. The main application of this detector is the Rn monitoring in buildings equipped with air conditioning and ventilation systems as well as Rn monitoring in special drilled wells for earthquake prediction (see Fig.10 and [23]for more details).

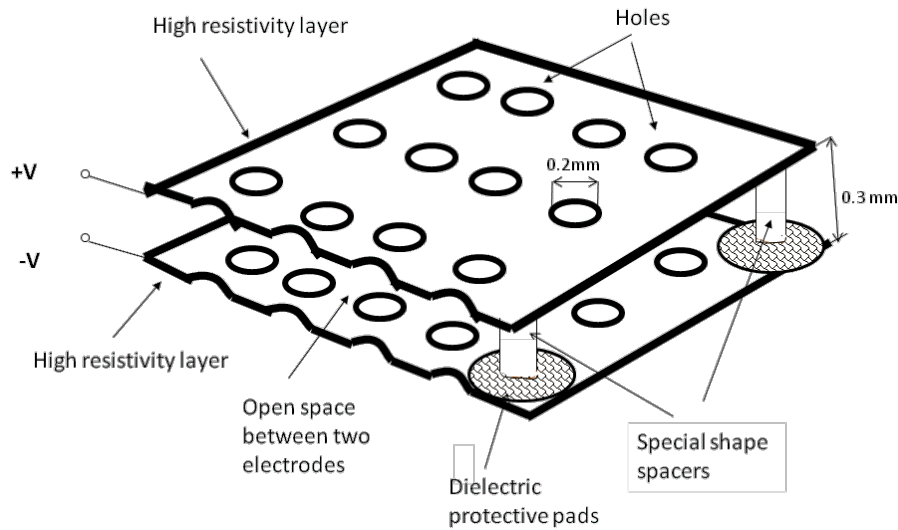

Fig.9. Schematic drawing of a wall-less detector with resistive electrodes

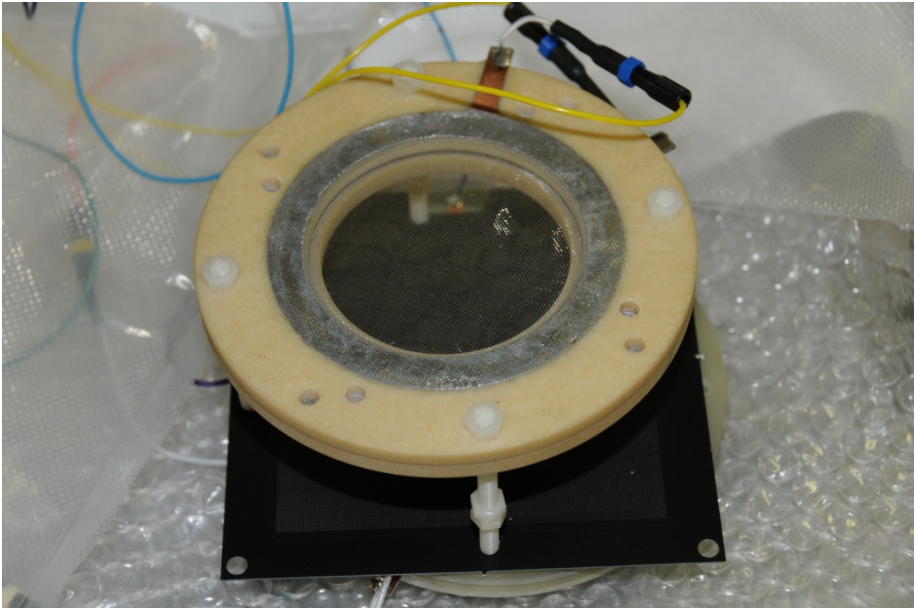

Fig.10. Photograph of one of our prototypes of the Rn detector r consisting of a drift mesh and the restive GEM with an active area of 10x10cm$^2$.

**4b. Super sensitive flame detectors**

In previous paper [24] a detector of flames having a sensitivity 1000 times higher than the best commercial detectors has beendescribed. It is a single-wire counter filled with TMAE vapors. It will be very attractive to develop a similar detector based on GEMs: this detector will have a compact planar geometry and the advantage to additionally increase the sensitivity by using large-detection areas.

Tests have, however, shown that standard GEMs or resistive GEMs exposed to TMAE vapors suffer from a leakage current across the hole's walls (as in the case of the humid air) while the wall-less GEMs feature a leakage current close to zero. Measurements performed with a wall-less GEM having an effective area of 100 cm$^2$ show that its sensitivity to flames is 4000 times higher than that of commercial sensors.

**5. Other interesting developments**

We have described, as examples, only three selected developments in advanced micropattern detectors with resistive electrodes made out of strips having readout electrodes under them. Many more interesting works are currently in progress, carried on not only by our team but by others as well. For example, in [10] a microgap-microstrip resistive plate chamber was described and, in a parallel work [25], it was shown that with a similar detector (actually, the strip pitch was larger), one can achieve simultaneously high time (~70 ps) and position resolutions (~50 μm). Breskin group develop resistive CAT [26] and this list can be continued. However, probably the most impressive is the development of large-area (~2 m$^2$) resistive strip MICROMEGAS for the ATLAS upgrade [5].

## 6. Conclusions

A new generation of micropattern gaseous detectors with resistive-strip electrodes combined with metallic 2D readout strips has been developed. It offers excellent position resolution and spark self-protection. In parallel a similar approach has been developed by the MAMMA collaboration and resistive strip MICROMEGAS will be employed in the ATLAS small wheel. More developments in the same direction are in progress. Of course, these detectors have limited rate capabilities and this can be an issue in high rate environment, although some improvements in their high rate performance are still possible.
The results obtained so far are very encouraging making us believe that new micropattern detectors with resistive electrodes may have very relevant applications in the future.